# The Medium Energy (ME) X-ray telescope onboard the *Insight*-HXMT astronomy satellite

Cao Xuelei[1*], Jiang Weichun[1††], Meng Bin[1††], Zhang Wanchang[1††], Luo Tao[1††], Yang Sheng[1††], Zhang Chunlei[1††], Gu Yudong[1††], Sun Liang[1††], Liu Xiaojing[1††], Yang Jiawei[1††], Li Xian[1††], Tan Ying[1††], Liu Shaozhen[1††], Du Yuanyuan[1††], Lu Fangjun[1††], Xu Yupeng[1,7††], Zhang Shuangnan[1,7], Wang Huanyu[1], Li Tipei[1,2,7], Zhang Chengmo[1], Wen Xiangyang[1], Ge Mingyu[1], Zhou Yupeng[3], Xiong Shaolin[1], Zhang Shu[1], Zhang Yongjie[1], Cheng Zehao[4], Zhang Fei[1], Li Maoshun[1], Liang Xiaohua[1], Gao Min[1], Yang Enbo[5], Liu Xiaohang[6], Liu Hongwei[1], Yang Yirong[1], Zhang Fan[1]

[1] *Key Laboratory for Particle Astrophysics, Institute of High Energy Physics, Chinese Academy of Sciences, Beijing 100049, China;*
[2] *Tsinghua University, Beijing 100084, China;*
[3] *Beijing Key Laboratory of Space Thermal Control Technology, Beijing Institute of Spacecraft System Engineering, Beijing 100094, China;*
[4] *University of Science and Technology of China, Hefei 230022, China;*
[5] *Guizhou University, Guiyang 550025, China;*
[6] *Yantai University, Yantai 264005, China.*
[7] *University of Chinese Academy of Sciences, Chinese Academy of Sciences, Beijing 100049, China*

The Medium Energy X-ray telescope (ME) is one of the three main telescopes on board the *Insight* Hard X-ray Modulation Telescope (*Insight*-HXMT) astronomy satellite. ME contains 1728 pixels of Si-PIN detectors sensitive in 5-30 keV with a total geometrical area of 952 $cm^2$. Application Specific Integrated Circuit (ASIC) chips, VA32TA6, is used to achieve low power consumption and low readout noise. The collimators define three kinds of field of views (FOVs) for the telescope, 1°×4°, 4°×4°, and blocked ones. Combination of such FOVs can be used to estimate the in-orbit X-ray and particle background components. The energy resolution of ME is ~3 keV at 17.8 keV (FWHM) and the time resolution is 255 μs. In this paper, we introduce the design and performance of ME.

*Corresponding author (email: caoxl@ihep.ac.cn)
††These authors contributed equally to this work.

## 1 Introduction

The Medium Energy X-ray telescope (ME), which is shown in Fig. 1-1, covering 5-30 keV, is one of the three main telescopes on board the Hard X-ray Modulation Telescope (HXMT) astronomy satellite, dubbed *Insight*-HXMT after its launch on June 15, 2017. The energy range of ME bridges between the energy bands of the high energy X-ray telescope (HE, about 20-250 keV) and the low energy X-ray telescope (LE, about 1-15 keV) of *Insight*-HXMT, so that the properties of many X-ray sources can be better constrained [1-3].

There are four pixels in one ceramic package, called Si-PIN detector hereafter. ME employs an array of 432 Si-PIN detectors. The whole array includes 1728 pixels with a geometrical area of 56.25 mm$^2$ each, and a total geometrical area of 952 cm$^2$. A similar technique was used by the hard X-ray detector (HXD) on board *Suzaku*, where 64 large active area Si-PIN detectors (pixel size: 21.5mm×21.5mm×2mm) were used to cover 10-70 keV.[4] HXD utilized Si-PIN diodes and high-Z GSO scintillators in anti-coincidence to achieve a lower background level[5]. Compared to the *Suzaku*/HXD, ME has a smaller pixel size but a much larger total geometrical area. To reduce the electronics power and the mass of ME, the front-end electronics and shaping electronics are realized by Application Specific Integrated Circuit (ASIC) chips. The data acquisition circuit is controlled by a Field Programmable Gate Array (FPGA).

In this paper, we describe the design, basic functions, and the on-ground test results of ME.

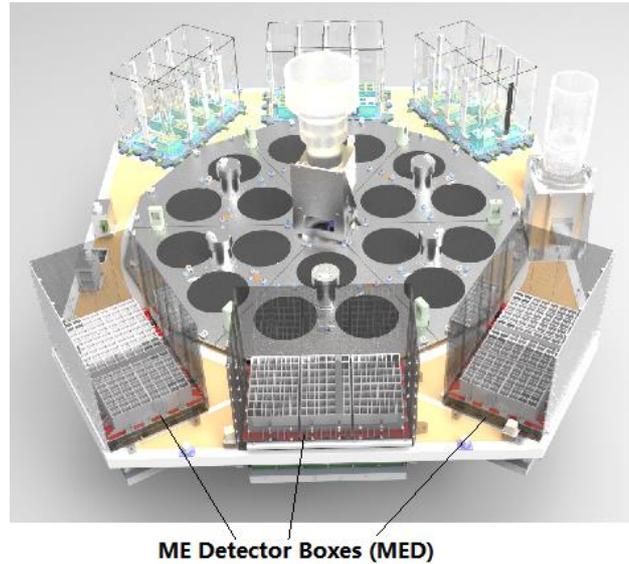

Figure 1-1 ME on the *Insight*-HXMT

## 2 Instrument description

### 2.1 Overview

ME consists of three medium energy detector boxes (MEDs) and one electric control box (MEB). Data collected by the three MEDs are sent to MEB through Low-Voltage Differential Signaling (LVDS) interfaces, while MEB communicates with the satellite platform with both LVDS and 1553B interfaces. The LVDS interface is mainly for the science data, and the 1553B interface is mainly for the engineering data and commands.

The main technical specifications of ME are summarized in Table 1.

**Table 1 Main features of ME**

| Items | Technical specifications |
|---|---|
| Energy range | 5-30 keV |
| Detector type | Si-PIN |
| Pixel size | 12.5 mm × 4.5 mm, 1 mm thick |

| Total geometrical area | 952 cm$^2$ |
|---|---|
| Energy resolution (FWHM) | ~3 keV at 20 keV |
| Time resolution | 255 μs |
| Detector working temperature | -50 ℃~-5 ℃ |
| Maximum count rate allowed | 2.2×10$^4$ cts/s |
| Power | 155 W |

ME can be divided into three main components, which are collimators, detectors and electronics. The collimators define three kinds of FOVs, 1°×4° (FWHM), 4°×4°(FWHM), and the blocked ones, which can be used to estimate different background components in orbit. Combination of these FOVs can also optimize the requirements for high-cadence scanning on the Galactic plane and the high sensitivity pointed observations. The Si-PIN detectors are developed by the Institute of High Energy Physics (IHEP) and are sensitive to X-ray photons in the energy range of 5-30 keV. The ME electronics utilizes ASIC chips to collect, amplify, digitize, and transmit the output signals of the detectors.

The functional block diagram of ME is shown in Fig. 2-1. The main functions of the MEDs are listed below:

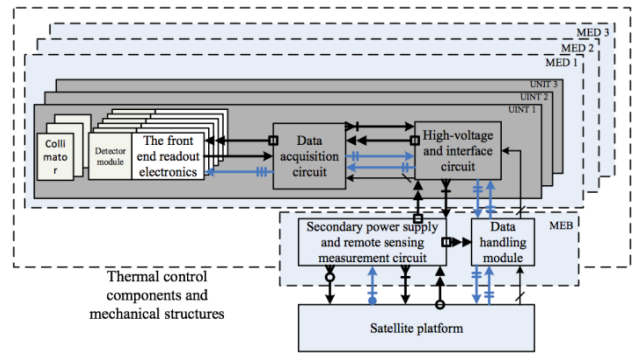

Figure 2-1 The functional block diagram of ME.

1) To detect X-ray photons, convert the photon energy and arrival time information into digital signals, and send them to MEB ;
2) To receive the control commands from MEB;
3) To generate the engineering data and transmit them to MEB;
4) To receive the clock and reset signals from MEB;
5) To transmit the analog and digital telemetry signals to MEB;

The main functions of the MEB are as follows:

1) To receive the science data and the engineering data from MEDs and then layout and frame them;
2) To generate data about the working conditions and the instruction link communication status of each module and to calculate the count rate of each module in MEDs;
3) To receive the remote-control commands from the satellite platform;
4) To forward the control commands to MEDs;
5) To transmit the telemetry data to the satellite platform;
6) To receive and forward the GPS pulse-per-second from the satellite platform;
7) To send the science data and the engineering data to the satellite platform;
8) To supply electronic power to MEDs.

## 2.2 Modular design

The ME system contains 432 Si-PIN detectors. Each Si-PIN detector has four pixels and therefore 1728 readout electronic channels are needed. We used 54 ASIC chips, each of which contains 32 channels, to meet the requirements of the readout system of all the Si-PIN detectors.

To improve the reliability of the whole system, a modular design is adopted, and each module can work independently. Each MED contains three identical detector units. All the detectors in a detector unit share a set of data acquisition modules and high voltage power supply module. In case any of the nine detector units fails, the rest can still work normally.

A detector unit is further divided into six detector modules. Each module has eight Si-PIN detectors, one ASIC chip, and the auxiliary circuit. Again, each module works independently and does not affect the other modules in case of failure. According to the function of the ASIC chips, the noise threshold of each channel is adjustable.

The diagram of modular design is shown in Figure 2-2.

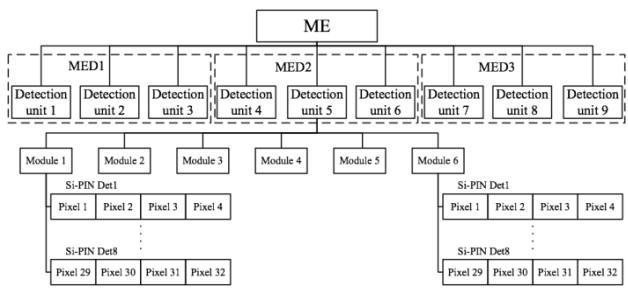

Figure 2-2 The architecture of ME.

### 2.3 The mechanical structure

The structure of one MED is shown in Figure 2-3.

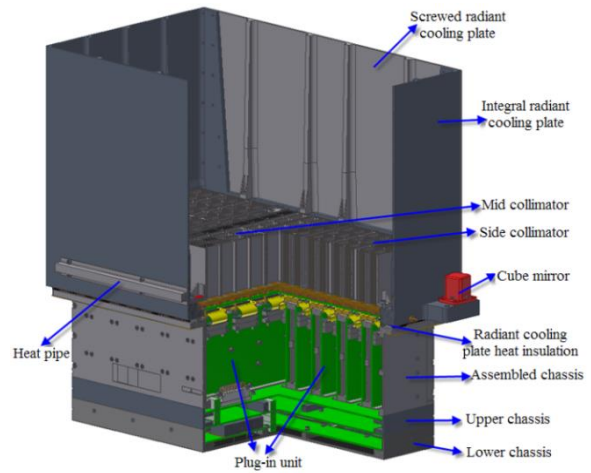

Figure 2-3 The structure of one ME detector box.

In each MED box, there are nine independent plug-in structures, and each plug-in structure contains two detector modules (see Figure 2-4).

The Si-PIN detectors are bonded to copper plates with heat conductive adhesive. The copper plate serves as a heat conductor to transmit the heat of the detector to the radiative surface to ensure the low working temperature of the Si-PIN detectors.

The electronics boards are fixed on the aluminum frame of the plug-in structure. To reduce the influence of the electronics on the detector temperature, thermal insulation between the aluminum frame and the copper plate is applied. The detector module and the electronics board are connected by a flexible cable (see Figure 2-5).

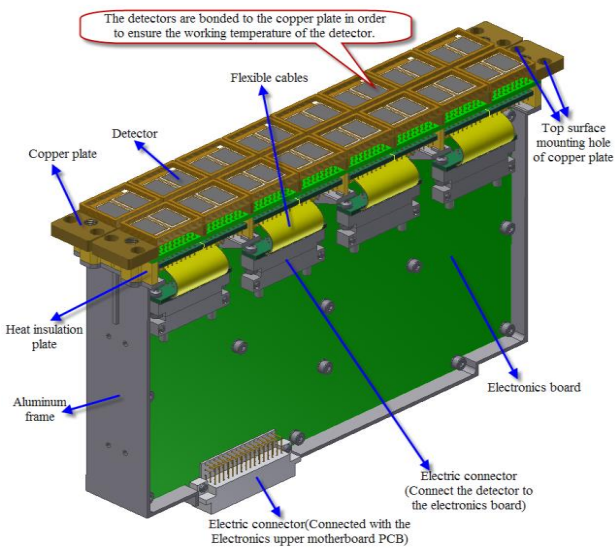

Figure 2-4 The structure of a plug-in structure.

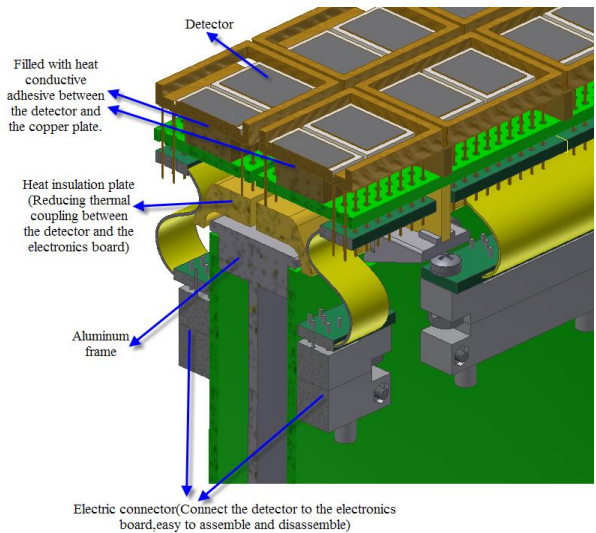

Figure 2-5 Thermal insulation in the plug-in structure and the flexible cable connection.

To ensure the working temperature of each detector, a copper plate and a radiative cold plate are directly connected with screws. In addition, the side of the plug-in structure is also connected with the structure of the MED by screws to ensure the mechanical strength of the plug-in structure and the overall structure of the MED. The detailed MED internal connection is shown in Figure 2-6.

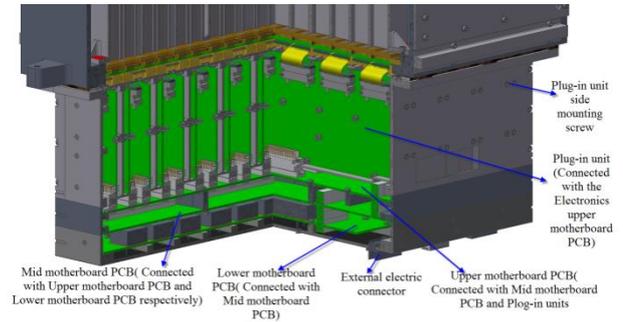

Figure 2-6 Internal structure of MED.

Figure 2-7 is a photo of the detectors of a plug-in structure. The photo of an MED (without the radiative cooling plate) is shown in Figure 2-8.

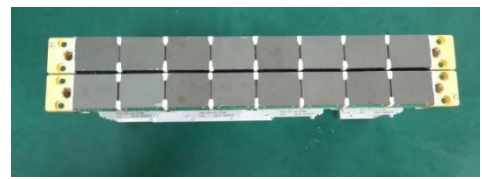

Figure 2-7 Photo of the detectors of one plug-in structure.

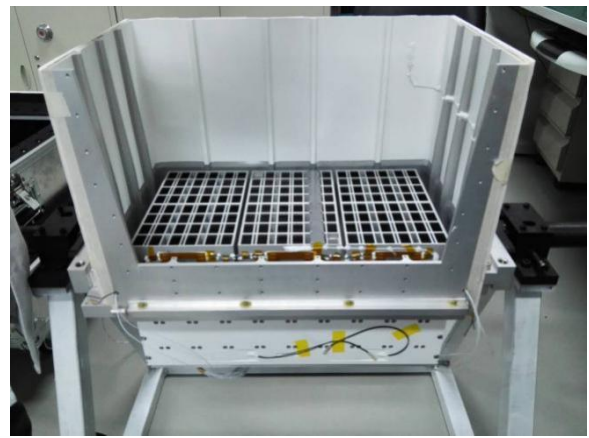

Figure 2-8 Photo of an ME detector box (without the radiative cooling plate).

**2.4 Si-PIN detector**

Ion-implanted planar technology is employed to develop the ME Si-PIN detectors. Different from the traditional PN junction, ion implantation method is used to form a high resistance intrinsic layer for this detector. Therefore, the leakage current is as low as several pA even with the depletion voltage applied to the electrodes.

The Si-PIN detector used by ME has a pixel size of 56.25 mm$^2$ (12.5 mm × 4.5 mm) and a thickness of 1 mm. One guard ring (red dash lines) is shared by two pixels (blue areas) on a silicon chip, as shown in Figure 2-9. Two silicon chips are packaged in a ceramic shell and thus four pixels compose one Si-PIN detector.

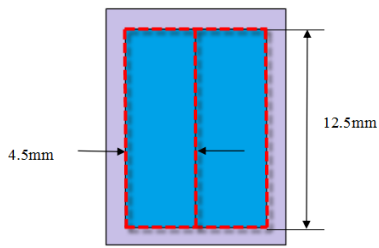

Figure 2-9 Block diagram of a silicon chip.

According to the pixel size of the design, the junction capacitance of the detector (proportional to the area and inversely proportional to the thickness) could be limited to ~ 5 pF. The surface passivation process and guard ring design can effectively minimize the surface leakage current and reduce the noise on the circuit. By grounding the guard ring, the leakage current is lower than 10 pA at -25℃ (the typical in-orbit working temperature of ME) with a depletion voltage of 150 V.

Figure 2-10 shows the leakage current of Si-PIN detectors. Its inset demonstrates the leakage current reaches ~20 pA at -22.5 ℃ and ~70pA at -5℃ with a bias voltage of 180 V.

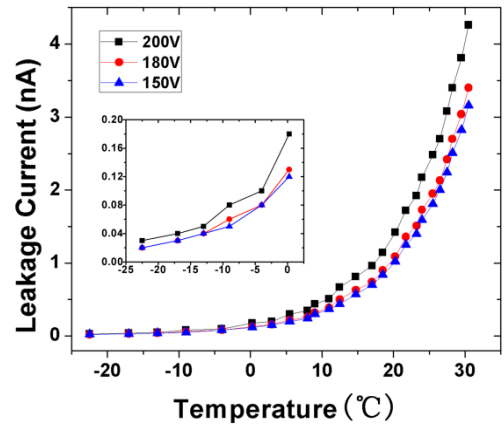

Figure 2-10 Leakage current of Si-PIN versus temperature.

A 50 μm beryllium (Be) window is pasted to the ceramic package to shield the optical light and protect the surface of the Si-PIN pixels. We use silver paste to act as adhesive for the Be plate to make it grounded and to ensure a stable inner electromagnetic environment. Be has advantages of low density, high strength, and high X-ray transmittance. 96% of the 5 keV X-rays can penetrate the Be window (Figure 2-11).

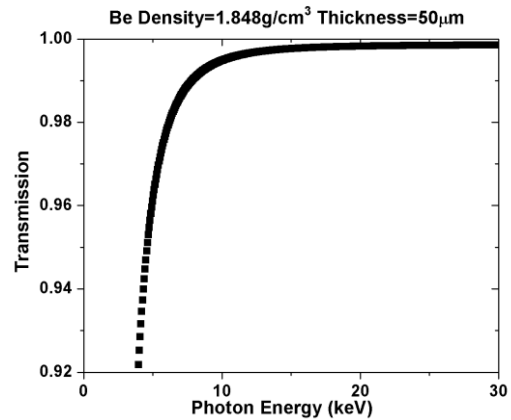

Figure 2-11 Transmittance curve of the 50 μm Be window.

**2.4 Read-out electronics**

ME contains 1728 Si-PIN pixels and thus 1728 channels of readout electronics are needed. Given the limited power and volume on the satellite, it is impossible to realize the readout electronics by discrete devices. Therefore, the front-end readout electronics are based on ASIC technology [6].

**2.4.1 Read-out electronics and the application of ASIC technology**

The ASIC used in MED is an improved version of VA32TA6 (Manufacturer: GM-Idea Company). It is a low power and rad-hard chip with 32 channels of low-noise charge sensitive amplifier (CSA)-shaper circuits. Each channel includes a slow shaper, a fast shaper, a discriminator, the trigger logic, and the sample & hold and ADC circuits (Figure 2-12)[7].

The energy resolution of ME is 3.0 keV (FWHM) at 20 keV. This energy resolution corresponds to 829 e− in ENC (equivalent noise charge). The noise includes detector noise and electronics noise, which can be affected by matching parameters of the detector and the ASIC. The energy range of ME is 5-30 keV, and the noise level needs to be very carefully controlled.

A CSA circuit is a current integrator that produces a voltage output proportional to the integrated value of the input detector current. The output of a CSA enters the corresponding slow shaper and fast shaper. The slow shaper forms a Gaussian signal with amplitude proportional to the photon energy. The shaping time of the slow shaper is 1 μs. The fast CR-RC (Capacitance Resistance – Resistance Capacitance) 300 ns shaper, with an adjustable slew rate, is followed by a level-sensitive discriminator. The discriminator is preceded by a high-pass filter (not shown in Figure 2-12) with a very low cut-off frequency to reduce the offset-spread across the chip.

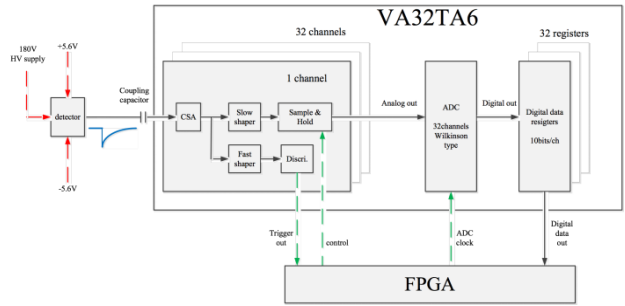

Figure 2-12 Block diagram of detector and read out electronics.

If the signal amplitude is higher than the threshold voltage of the discriminator, a trigger pulse will be generated. FPGA receives the trigger pulse, records a valid case, and then enables the sample & hold function of the ASIC after a delay of nearly 1 μs. The above sequence is shown in Figure 2-13.

The digitization of the sampled output of the slow shaper is accomplished by starting a digital counter coincident with a linear analog voltage ramp, starting at the voltage $V_{ref}$. The ramp voltage is compared to the DC signal value in each channel using an analog discriminator. The discriminator will then fire with a time delay proportional to the signal value. This discriminator signal will lock the digital counter value in a register for each channel. The ADC used in MED is the Wilkinson type 10-bit ADC, which can digitize all channels in parallel.

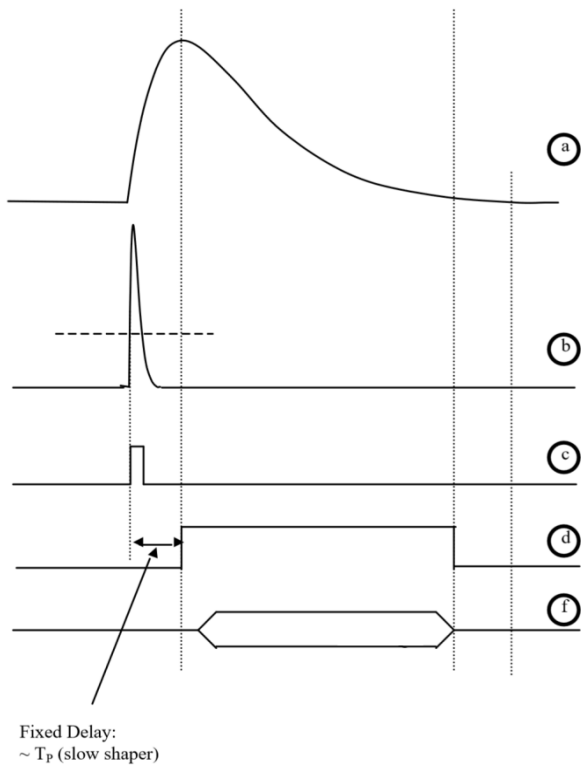

Figure 2-13 Signal, trigger, and data sequences in the ASIC of ME.

(a. slow shape, b. fast shape, c. trigger out, d. sample and hold control, f. ADC clock.)

**2.4.2 Data acquisition module**

There is one data acquisition module for each detector unit. The main components of the data acquisition module are shown in Figure 2-14. The main processing unit is an FPGA. The data acquisition module communicates with the data handing module through the LVDS interface. Every six ASICs in one detector units are configured and controlled by one FPGA. The configuration bits for the six ASICs are stored in an EEPROM. The engineering data are buffered in an SRAM. The power monitor can generate the power when a reset signal is sent. The FPGA can be reset by telecommand as well. The FPGA generates one bit digital telemetry to indicate the work status. There are two HV supply modules in one detector. They act as the cold backup for each other. The HV supply modules can be turned on or turn off by the FPGA. The FPGA also monitors if there is single event latchup (SEL) of ASICs and control the low dropout regulator (LDO) to repower the ASICs in order to relieve the latchup. The serial ADC can convert the analog temperature telemetry signals of ASICs and detectors to digital engineering data. The FIFO is the scientific data buffer. The serial DAC can generate the threshold for determining if there is SEL.

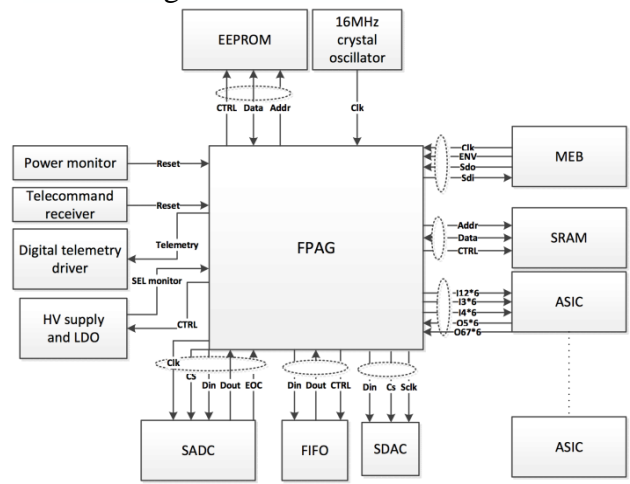

Figure 2-14 block diagram of ME data acquisition module in MED.

**2.4.3 Data handling module**

The data-handling module is responsible for communicating with the satellite platform and completing data exchange and control between MED and the satellite platform. The main functions of the data handling module are as follows: (1) to read the science data and the engineering data from MED and then to upload them to the satellite platform through the LVDS interface and 1553B bus interface, respectively; (2) to receive the Coordinated Universal Time (UTC) code broadcasted by the satellite platform and then transfer them to MED via LVDS interface; and (3) to receive the data injection commands from the satellite platform via 1553B, to process and send them to the appropriate MED for implementation. The satellite platform provides GPS pulse-per-second and 5 MHz clock pulses. The data-handling module also receives and transmits these pulses. The working-state monitoring parameters of the three MEDs are aggregated in the data handling module and then sent to the satellite platform. The related interfaces of the data handling module are shown in Figure 2-15. Two identical circuit boards in the data-handling module are connected to each other through four 74-pin sockets. They act as the backup for each other.

The data handling module contains one single-chip microcomputer (SCM) and one FPGA as the main processing unit. The SCM controls the 1553B protocol chip J61580R to communicate with the satellite platform via 1553B Bus. Instructions from the platform, including data injection and the UTC time code, are received by the SCM and stored into the SRAM, and the FPGA will read them out and send them to the corresponding MEDs by LVDS. The FPGA receives engineering data and writes them into the SRAM, and the SCM reads them out and sends them to the platform via 1553B Bus. The science data from the three MEDs are received by the FPGA and are sent to the platform through the LVDS interface.

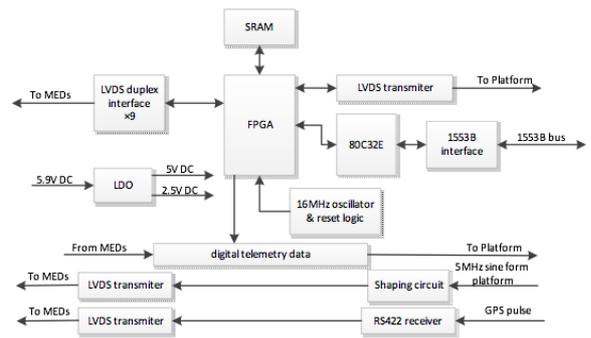

Figure 2-15 Block diagram of ME data handling module.

## 2.5 Collimators

The collimators are mounted on top of the plug-in structures to limit the FOVs of the telescope. There are three kinds of FOVs, i.e., $1°\times4°$(FWHM), $4°\times4°$(FWHM), and the blocked ones, which can be used to estimate the internal and particle components of the in-orbit background. Among the eighteen detector units, fifteen units have an FOV of $1°\times4°$, two have an FOV of $4°\times4°$, and one is blocked.

The ME collimator adopts an aluminum alloy frame with tantalum foils inserted inside. The advantage of this scheme is that the collimation hole is completely made up of the tantalum foils and thus can shield the X-rays up to 30 keV.

The collimator in each MED is divided into three modules, in order to be manufactured more easily. The photo of three modules assembled in the MED is shown in Figure 2-16. The FOVs of the two side modules are $1°\times4°$ and the FOVs of the middle modules include $1°\times4°$, $4°\times4°$, and the blocked ones. Two in-orbit calibration radioactive sources are installed and each source blocks one detector, as noted in Figure 2-16. The number of detectors corresponding to each FOV is shown in Table 2.

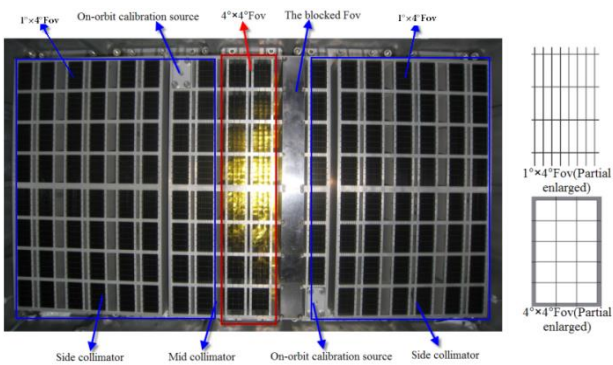

Figure 2-16 The top view of an MED showing the three collimator modules.

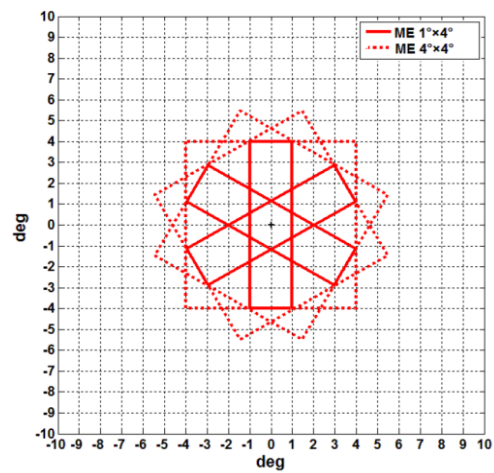

Table 2 The number of detectors and pixels that are employed

| FOV | The number of detectors | The number of pixels |
| --- | --- | --- |
| 1°×4° | 118 | 472 |
| 1°×4° | 16 | 64 |
| Blocked | 8 | 32 |
| Calibration sources | 2 | 8 |

2.6 Thermal Design

1) ME Telescope is mounted on the plate-Z side of the main structure of the load, directly exposed to the stars. The thermal control design of thermal insulation and passive radiation is adopted: 1) The heat consumption of MED is at 3 PCBs (17.3 W) in the bottom and 18 detector plug-in boards (12.78 W). In order to ensure the low temperature requirement of the medium energy detector, a 5 mm thick glass fiber reinforced plastic was designed between the X side of MED and the radiant cooling plate. The 10 mm thick amanium is designed to be placed between the X side of the detector plug-in aluminum frame and the plug-in PCB board.

2) The shell of MED is used as the radiative cooling surface and the thermal control coating is the SR107-ZK white paint coating. The radiative refrigeration plate with the structure of the aluminum plate installed in the detector is designed. Considering the sensitivity of the external heat flux and the eases of coating, on the outer side of the radiating surface of the radiating plate, the OSR coating with less degradation at the end and which is less affected by earth infrared radiation is selected, and SR-107ZK white paint coating is selected on the inside of the radiating surface of the radiating plate.

3) Due to the requirement of temperature uniformity being less than 5 ℃, two outer heat pipes were designed at the bottom of both sides of the long end of the medium energy radiant cooling plate.

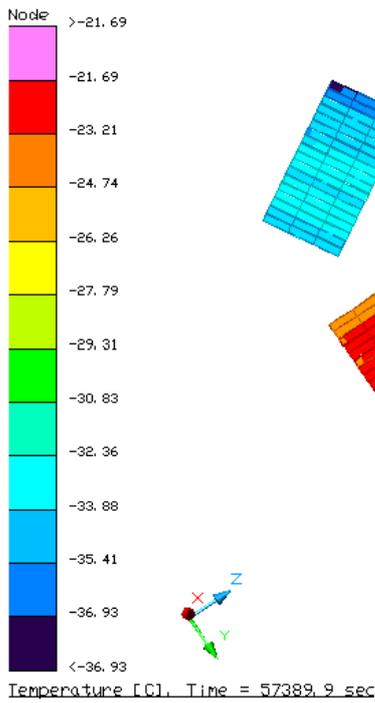

Figure 2-17 Temperature distribution of MEDs under high temperature condition

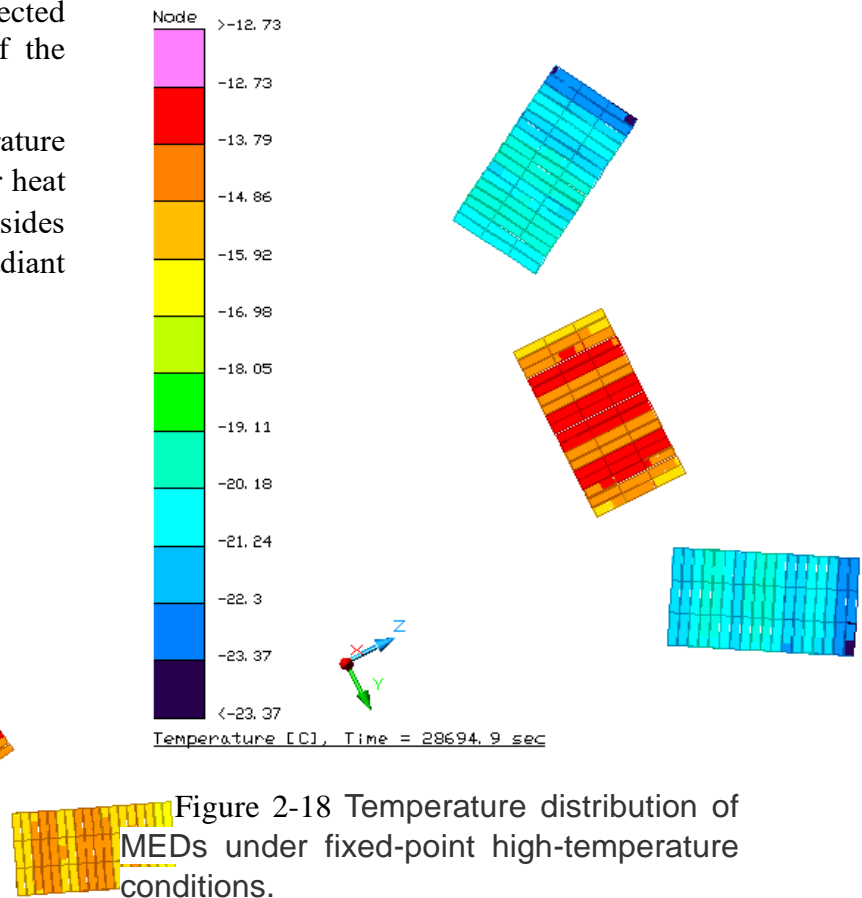

Figure 2-18 Temperature distribution of MEDs under fixed-point high-temperature conditions.

The temperature level of the medium energy detector crystal array is -42.3 ℃ ~ -19.3 ℃ (-50 ℃ ~ -10 ℃), and the maximum temperature difference of a single detector crystal is 3.4 ℃ (< 5 ℃). The temperature level of the shell of MED is -37.8 ℃ ~ -14.9 ℃ (-40 ℃ ~ 25 ℃). The temperature level of the medium energy detector crystal array is -33.4 ℃ ~ -9.1 ℃ (-50 ℃ ~ -5 ℃) and the maximum temperature difference of a single detector crystal is 3.4 ℃ (< 5 ℃) in fixed-point mode. The temperature level of the shell of MED is from -29.7 ℃ to -7.6 ℃ (-40 ℃ ~ 25 ℃), which can meet the corresponding temperature requirements. This demonstrates that the thermal design of the medium-energy detector is reasonable and feasible.

**3 Calibration and performance of the instruments**

A series of experiments and tests has been carried out for ME, including the calibrations, environment tests, as well as some function and interface tests. The energy response, temperature response, and dead time of ME are calibrated with radioactive sources and a double-crystal monochromator. The radioactive sources are $^{241}$Am and $^{57}$Co. Two pixels of the Si-PIN detector with similar readout electronics were tested on the PANTER X-ray test facility of the Max Planck Institute for Extraterrestrial Physics in 2013. The calibration of the ME qualification model was performed with a facility similar to PANTER, which is at IHEP of the Chinese Academy of Sciences (CAS).

**3.1 Calibration with radioactive sources**

All ME detectors were calibrated with integrated radioactive sources, $^{241}$Am or $^{57}$Co. To achieve a relatively uniform count rate distribution, the structure of the integrated radioactive sources is designed as in Figure 3-1. The structure consists of four single radioactive sources with an activity of 30 μCi each.

The responses of the detectors to the X-ray emission of $^{241}$Am (with emission lines at 13.9 keV, 17.8 keV, 21.6 keV, and 26.3 keV) and $^{57}$Co (with an emission line at 14.4 keV) were measured at different temperatures (-31℃~-5℃). The energy-channel (E-C) relations at different temperatures can be calibrated by these measurements.

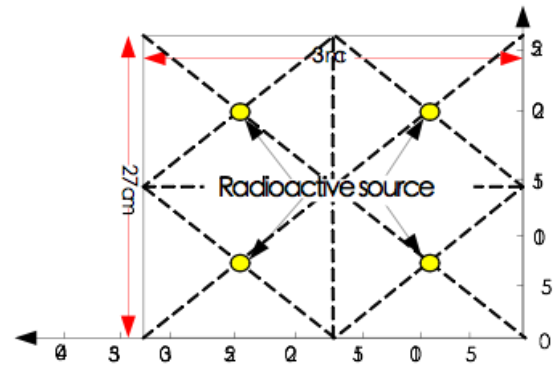

Figure 3-1 The structure of the integrated radioactive source ($^{241}$Am or $^{57}$Co).

**3.2 Calibration with the double-crystal X-ray monochromator**

The energy response and quantum efficiency calibration were implemented using the monochromatic X-ray beam on the calibration facility at IHEP, which was designed just for the ground calibration of ME and LE[9]. The calibration facility includes an X-ray tube, a double-crystal monochromator, and a vacuum chamber in which the detector box can be calibrated. The major characteristics of the calibration facility are shown in Table 3-1.

Table 3-1 Major characteristics of the calibration facility

| Characteristics | Value/range |
|---|---|
| Energy | 0.929~30.23 keV |
| Fraction of the monochromatic emission | 88%-99% |
| Beam flux | 0.1-580 photons/mm$^2$/s |

| Beam size | 40 mm×20 mm at 8 keV |
|---|---|
| Monochromaticity | 84 eV at 8 keV |
| Temperature Stability | 0.5 K in 193 k- 273 K . |

### 3.3 Calibration results

The calibration results of ME shown here are mostly performed with radioactive sources and the double-crystal X-ray monochromator at IHEP. The quantum efficiency (QE) results were obtained on PANTER.

**3.3.1 Analysis of energy spectra**

The response function of the 1728 Si-PIN detector pixels are calibrated with radioactive sources that contain a few emission lines, and four pixels are calibrated using the double-crystal X-ray monochromator. To simplify the analysis, we construct a function $F(c)$ that includes several Gaussian components to fit the measured spectra. The number of components $N$ is the same as the number of lines of the radioactive sources,

$$F(c) = \sum_{i=1}^{N} n_i \exp(-0.5(\frac{c - c_i}{\sigma_i})^2),$$

where $c$ is the channel of the spectrum (from 0 to 1023 here), $n_i$, $c_i$, and $\sigma_i$ are the intensity, central channel, and standard deviation of the $i$-th component (the line of the energy $E_i$), respectively. The energy resolution (FWHM) at energy $E_i$ is defined as $2.355\sigma_i$.

Figure 3-3 is a typical energy spectrum of $^{241}$Am measured by ME. In this figure the central energies of the Gaussian profiles are 10.56, 13.88, 17.73, 21.6, and 26.3 keV, which are the energies of the emission lines.

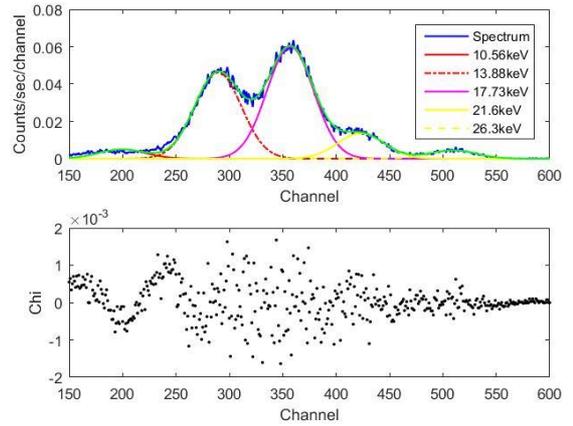

Figure 3-3 A typical energy spectrum of Am-241 fitted with a few Gaussian functions.

Figure 3-4 is a typical energy spectrum of $^{57}$Co. In this figure the central energy of the profile is 14.4 keV.

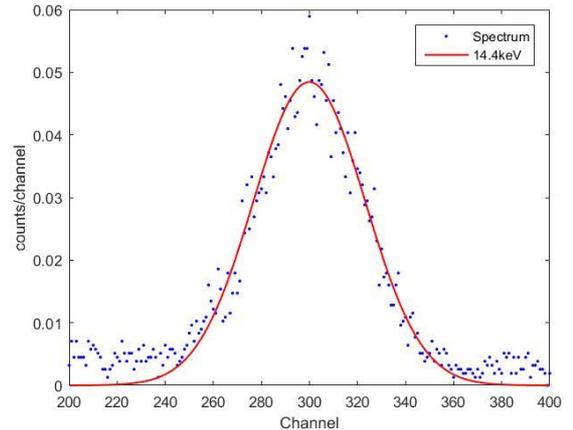

Figure 3-4 A typical energy spectrum of $^{57}$Co fitted with two Gaussian functions.

Figure 3-5 shows a typical energy spectrum using the double-crystal monochromator X-ray, and the central energy is 12 keV.

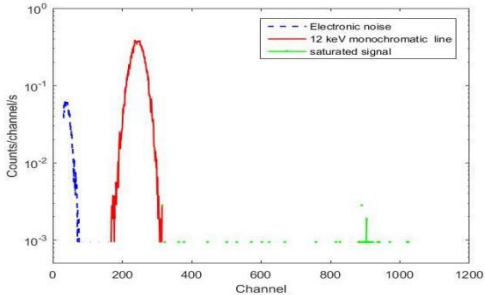

Figure 3-5 The measured spectrum of a 12 keV monochromatic line obtained by the ME qualification model. The 12 keV line is produced by a double-crystal monochromator.

The energy spectra of $^{241}$Am and $^{57}$Co measured by the 576 Si-PIN pixels in one MED are shown in Figures 3-6 and 3-7, respectively. In these two figures, the horizontal axis is the energy channel and the vertical axis represents the number of different pixels. The color represents the count rate per channel in logarithmic scale. The dead or non-working pixels are clearly shown in blue lines.

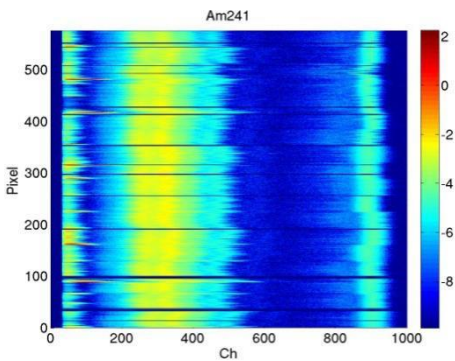

Figure 3-6 The energy spectra of $^{241}$Am measured by the 576 pixels in one MED (-25℃).

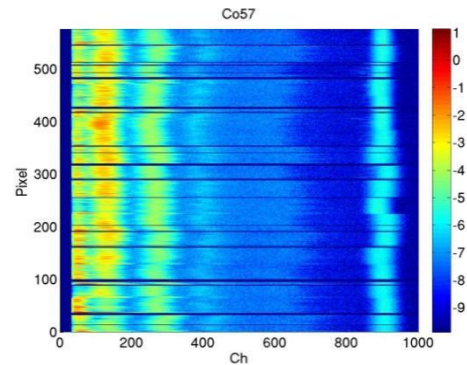

Figure 3-7 The energy spectra of Co-57 measured by the pixels in one MED (-25℃).

### 3.3.2 E-C relationship

Using $c_i$ and $E_i$, we can further obtain the energy to channel (E-C) relations. The E-C relations of one pixel at temperatures from -30℃ to 0℃ with a step of 5℃ are given in Figure 3-8, which shows that the linearity is quite good and the signal amplitude increases with temperature as expected. Figure 3-9 is the fitted parameters of the E-C relations of 576 Si-PIN pixels in one MED.

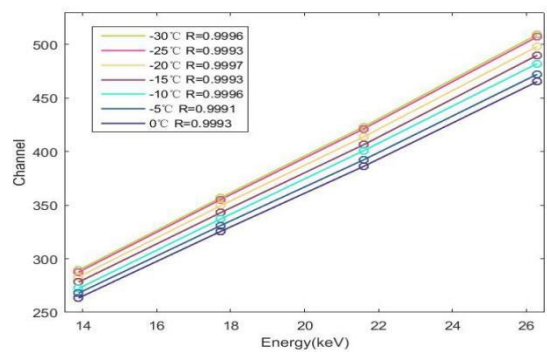

Figure 3-8 The E-C relations of one Si-PIN pixel at temperatures from -30℃ to 0℃ with a step of 5℃ (from top to the bottom).

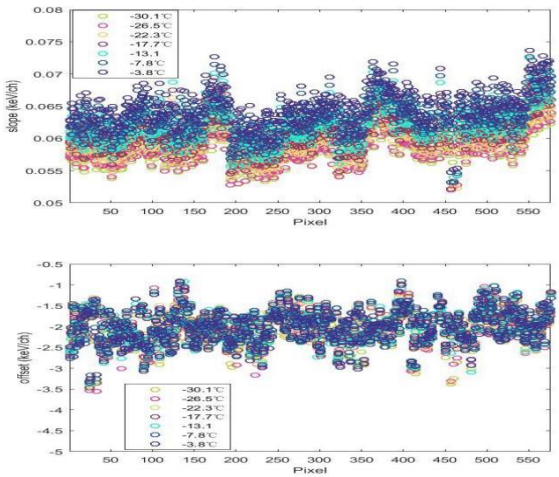

Figure 3-9 The parameters (intercept and slope) of a linear fitting to the E-C relations of the 576 Si-PIN pixels in one MED with the temperature between -30℃ and 0℃.

The values of slope and offset of the 576 Si-PIN pixels in one MED at different temperatures are shown in Figures 3-10 and 3-11 respectively. The slope increases as the temperature increases and the offset remains relatively stable with different temperatures. The signal amplitude increases by 0.6% per degree.

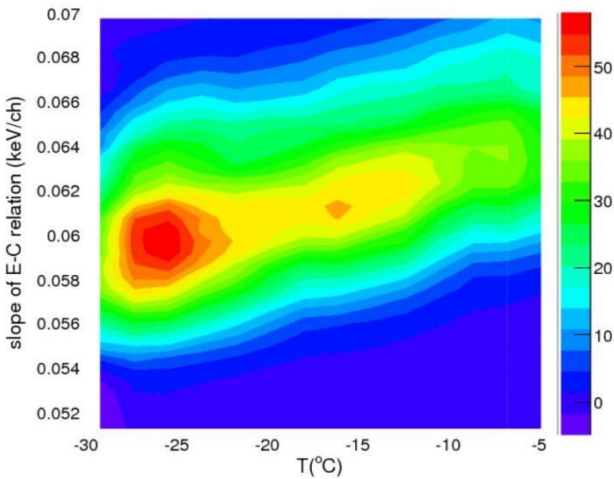

Figure 3-10 Distribution of the slope of the E-C relation for the 576 Si-PIN pixels in one MED at different temperatures. The slope of the E-C relation falls in an interval of 0.01. The colors from blue to red represent the pixel numbers.

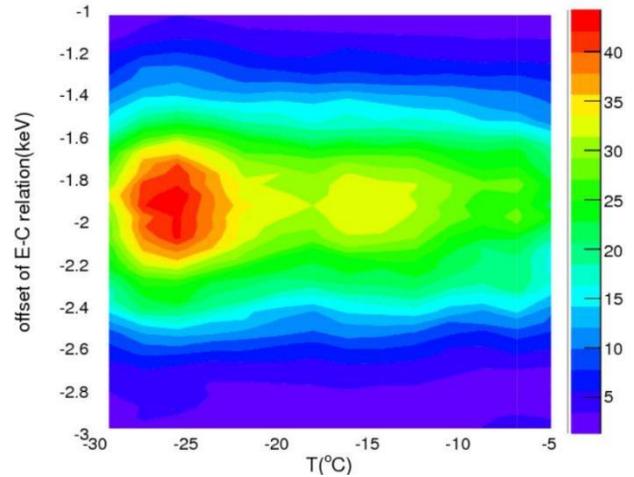

Figure 3-11 Distribution of the offset of the E-C relation for the 576 Si-PIN pixels in one MED at different temperatures. The offset of the E-C relation falls in between [-2.6, -1.2]. The colors from blue to red represent the pixel numbers.

### 3.3.3 Energy Resolution-E relationship

The energy resolution of ME is mainly determined by the noise of the detector and readout electronics. Because the resolution is dominated by the performance of the ASIC, which has smaller uncertainty when the input signal becomes larger, the energy resolution (FWHM) is better with increasing X-ray energy (see Figure 3-12).

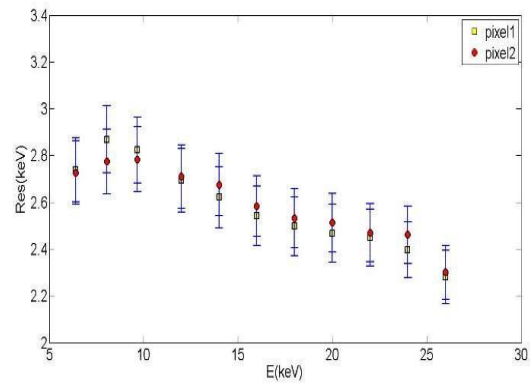

Figure 3-12 FWHM of two pixels in one MED at different energies.

Since the leakage current of Si-PIN detectors improves at lower temperatures, the energy resolution also improves at lower temperatures as expected (Figure 3-13).

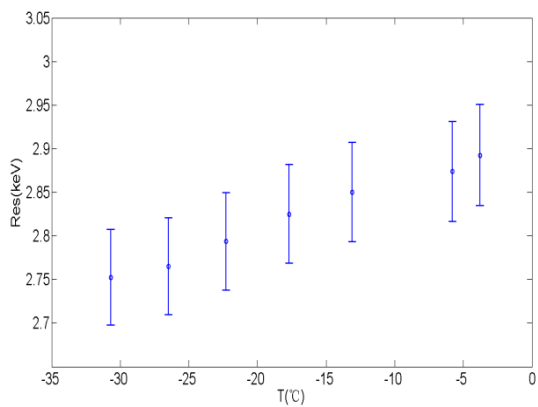

Figure 3-13 Change of FWHM at 17.73 keV with temperature.

**3.3.4 Quantum efficiency**

The quantum efficiency was obtained by the experiment carried out on the PANTER facility. In this experiment, the standard detector is the monitor of the beam line, which is a CCD type detector whose quantum efficiency had been calibrated [8]. The efficiency below 10 keV is limited by the ASIC threshold, shown in Figure 3-14. Using the results of the PANTER experiment, we simulated the quantum efficiency of ME in 5 to 40 keV, as shown in Figure 3-15.

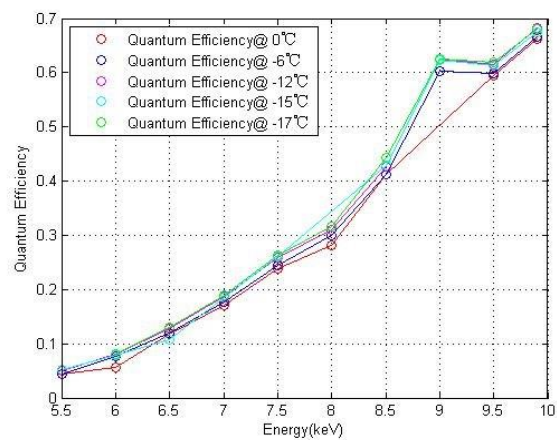

Figure 3-14 The quantum efficiency (QE) of the calibration module at different temperatures.

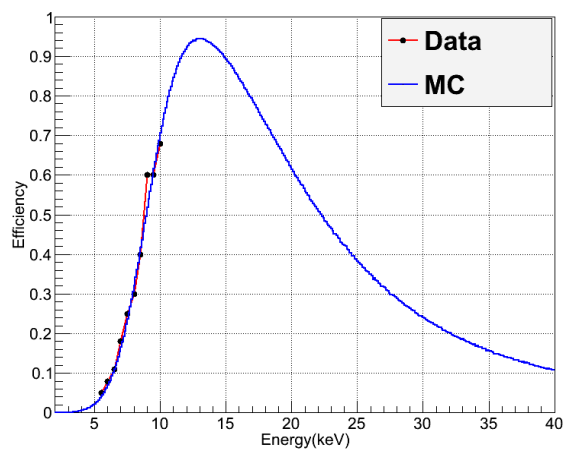

Figure 3-15 Simulated quantum efficiency of ME. The data points are obtained from the experiment on PANTER.

In the simulation model, the X-ray entrance window is the 50 μm thick Be window, whose transmission curve is shown in Figure 2-11. The detector model is a 1 mm thick Si depletion layer with a 0.8 μm thick Al front electrode.

For the energy range below 10 keV, the quantum efficiency changes with the values of the internal parameters of the ASIC chips, such as VTHR. However, these parameters cannot change the quantum efficiency much above 10 keV. Therefore, the 10-40 keV quantum efficiency is obtained by simulation.

### 3.3.5 Dead time

The dead time ($T_D$) of ME is a combination of ADC dead time ($T_{ADC}$) and readout dead time ($T_{RD}$). If *n* pieces of ASICs (out of the six ASICs in one unit that are controlled by one FPGA) are triggered in the same FPGA sampling period (16 μs), they are considered to be triggered simultaneously. The dead time is thus $T_D = T_{ADC} + n \times T_{RD}$. The typical values for TADC and TRD are 162.1 μs and 93.7 μs, respectively. The in-orbit count rate of ME is usually very low and thus only one ASIC is triggered in most cases. As a result, the arrival time interval spectrum is a standard negative exponential curve (Figure 3-16). The x-axis is in units of 6 μs, i.e., the precision of the time code. In case of a single ASIC trigger, the dead time is about 42 or 43 time code units, corresponding to 252~258 μs.

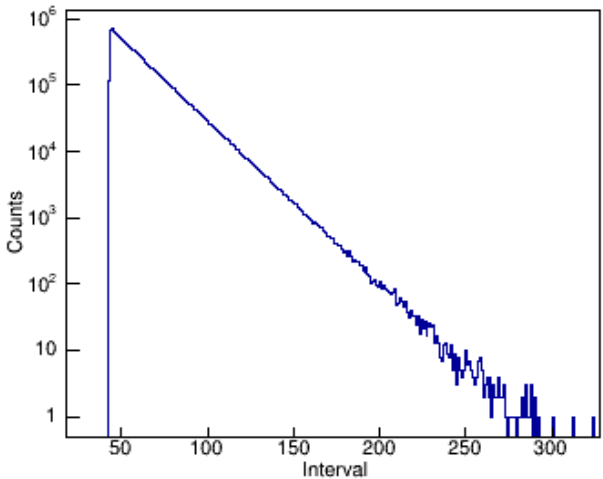

Figure 3-16 The arrival time interval spectrum of an MED obtained in the thermal vacuum test (-20℃). The horizontal axis is in units of 6 μs.

When the ME detectors work at high temperature (above 0℃) or observe a bursting source, the count rate can become too high and more than one ASIC in a unit will be triggered. The arrival time interval spectrum then is no longer a pure negative exponential curve. As shown in Figure 3-17, the dead time of the single-ASIC event is still 252~258 μs; however, if more than one ASIC chip is triggered, the dead time will increase by 90~96 μs. If six ASIC chips in a unit are triggered simultaneously, the dead time of ME is about 700~738 μs

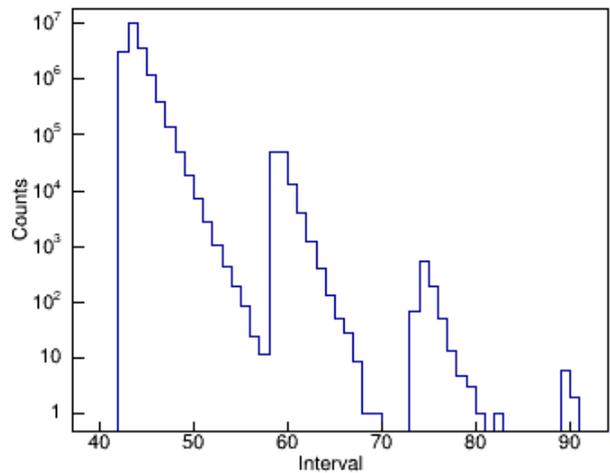

Figure 3-17 The arrival time interval spectrum of MED with high count rates . The horizontal axis is in units of 6 μs.

### 3.3.6 Relative time precision

The relative time precision was determined by jitters in charge collection time of Si-PIN detectors, the jitters in pulse triggering of ASIC, the jitters in propagation delay of electronics in MED, and the quantified precision of digital electronics in MED. The quantified precision is determined by the least significant bit of timing code. The collection time in Si-PIN is no more than 20 ns, the jitters in pulse triggering are no more than 300 ns, and the jitters in propagation delay of electronics are no more than 100 ns. The least significant bit of timing code corresponds to 6 μs. Consequently, the relative time precision of ME was mainly determined by the least significant bit of timing code, which was less than the technical requirement of 10 μs.

**4 Conclusions**

With an array of 1728 Si-PIN detector pixels, ME has achieved a geometrical area of 952 cm$^2$. The energy resolution is about 3 keV in the energy range of 5-30 keV and the dead time is about 255 μs. Before the launch on June 15, 2017, all flight products were calibrated. The energy response functions of the detectors at different temperatures were calibrated with radioactive sources and monochromatic energy X-ray beams. The energy resolution improves when the working temperature decreases and at the same time the signal amplitude increases by 0.6 % per degree.

*The authors express their thanks to the people helping with this work, and acknowledges the valuable suggestions from the peer reviewers. This work was supported by the Strategic Priority Research Program on Space Science, the Chinese Academy of Sciences, Grant No. XDA040102.*

Abbreviations:

ME---Medium Energy X-ray telescope
HXMT--- Hard X-ray Modulation Telescope
ASIC --- Application Specific Integrated Circuit
FOVs --- Fields of view
FWHM --- Full Width at Half Maximum
HXD --- Hard X-ray Detector
FPGA --- Field - Programmable Gate Array
MED --- ME Detector Box
MEB --- ME Electric Control Box
LVDS --- Low-Voltage Differential Signaling
CSA --- Charge Sensitive Amplifier
CR-RC --- Capacitance Resistance – Resistance Capacitance
UTC --- Coordinated Universal Time
SCM --- Single Chip System Module
RAM --- Random-Access Memory
ADC --- Analog to Digital Conversion
E-C --- Energy-Channel
QE --- Quantum Efficiency